\documentstyle[12pt]{article}
\newcommand{\p}[1]{(\ref{#1})}
\newcommand{\cp}{\mbox{$\cal P$}}
\newcommand{\e}{\eta}
\newcommand{\be}{\begin{equation}}
\newcommand{\bea}{\begin{eqnarray}}
\newcommand{\ee}{\end{equation}}
\newcommand{\eea}{\end{eqnarray}}

\begin{document}
\setcounter{page}0
\renewcommand{\thefootnote}{\fnsymbol{footnote}}
\thispagestyle{empty}
{\hfill  Preprint JINR E2-2000-11}\vspace{1.5cm} \\
\begin{center}
{\large\bf
Auxiliary representations of Lie  algebras and the BRST constructions.
}\vspace{0.5cm} \\

{\v{C}estm\'{\i}r Burd\'{\i}k}\footnote{E-mail: burdik@dec1.fjfi.cvut.cz}
\vspace{0.5cm} \\

{Department of Mathematics,
Czech Technical University,\\
Trojanova 13, 120 00 Prague 2}\vspace{0.5cm} \\

A. Pashnev\footnote{E-mail: pashnev@thsun1.jinr.dubna.su}
and M. Tsulaia\footnote{E-mail: tsulaia@thsun1.jinr.dubna.su}
\vspace{0.5cm} \\
{\it JINR--Bogoliubov Theoretical Laboratory,         \\
141980 Dubna, Moscow Region, Russia} \vspace{1.5cm} \\
{\bf Abstract}
\end{center}
\vspace{1cm}

The method  of construction of auxiliary representations for a given
Lie algebra is discussed in the framework of the BRST approach.
The corresponding BRST charge turns out to be
non -- hermitian.   This problem is solved by the introduction
 of the additional kernel operator in the definition
of the scalar product in the Fock space. 
The existence of the kernel operator is proven for any Lie algebra.

\begin{center}
{\it Submitted to Modern Physics Letters}
\end{center}

\newpage\renewcommand{\thefootnote}{\arabic{footnote}}
\setcounter{footnote}0\setcounter{equation}0
\section{Introduction}

The  BRST quantization procedure  for a  system of the first
class constraints \cite{FV} -- \cite{BV} 
 is straightforward. By the definition, the
first class constraints
form a closed algebra with respect to  the commutators
(the Poisson brackets).
For simplicity we  consider only linear algebras -- Lie algebras
of constraints.

More general systems  include the second class constraints as well,
whose commutators contain terms which are nonzero on mass shell
(on the subspace where all constraints vanish).
 In the simplest
cases  these terms are a numbers or central charges, but sometimes,
they are operators which act
nontrivially on the space of the physical states.
Moreover, the commutators between these
 operators and the constraints can be nontrivial. In some cases
the total system of the constraints and the
operators mentioned above form a Lie algebra.

So, in such cases we have a system of operators which form a Lie algebra,
but the physical meaning of different operators is different. Some of them
play the role of constraints and annihilate the physical states, others
are nonzero and simply transform the physical states into other ones.
It means, that in the BRST approach for the description of the
corresponding physical system we can not use the standard BRST charge
for the given Lie algebra.
Instead, we have to construct the nilpotent BRST
charge in  a manner, that
some of the operators play the role of the first class constraints,
others are second class constraints and the others
do not  imply any conditions on the physical
space of the system.

In this letter we demonstrate the possibility of a different
BRST constructions for the given system of generators,
which form a given Lie
algebra and have different physical meaning.

In the Section 2 we discuss the general method of the BRST
quantization, when some of the Cartan generators are excluded
from the total system of constraints. The method is based on the 
introduction of an auxiliary representations of the  algebra
under consideration,
in the way that these representations
 effectively lead to the desired properties of
initial generators: some of them are of the first class,
some - of the second class and rest of the generators 
(only  the Cartan ones)
do not imply any equations on the space of physical  states.
The auxiliary representations of the algebra are constructed
by means of some additional degrees of freedom, which are 
usually exploited  for  the quantization of the systems with the 
second class constraints \cite{FS} -- \cite{EM}.

In the Section 3 we describe the construction of auxiliary representations
of the algebra  by means of the method of induced representations.
The resulting operators act in a new  Fock space,  generated
by a set of additional creation and annihilation operators.
The construction automatically destroys the hermiticity properties
of the generators. The consequence of this fact is that the BRST
charge ${\cal Q}$ becomes non -- hermitian: ${\cal Q}^+$ 
is not equal to ${\cal Q}$. 
As an example of this general construction
the case of the $so(3,2)$ algebra is considered in details.

In the Section 4  we show, how these hermiticity properties are restored
by introduction of a new scalar product with some kernel $K$ in this
auxiliary  Fock space. As a consequence, the new operator $K {\cal Q}$ 
becomes hermitian and can be used for the construction of 
the lagrangians, which are
gauge invariant due to the nilpotency of (non -- hermitian) BRST 
charge ${\cal Q}$.

\setcounter{equation}0\section{The general method}
 In this section, we  describe the
method of the BRST construction, which leads to the desirable division
of the generators of a given Lie algebra
 into the first and second class constraints.
Let ${\hat H}^i,\;(i=1,...,k)$ and ${\hat E}^{\alpha}$
be the Cartan generators and root vectors
of the algebra with the following commutation relations
\begin{eqnarray}
\label{commutator}
&&\left[{\hat H}^i,{\hat E}^\alpha\right]=\alpha(i) {\hat E}^\alpha,\\
&&\left[{\hat E}^\alpha,{\hat E}^{-\alpha}\right]=\alpha^i {\hat H}^i,\\
&&\left[{\hat E}^\alpha,{\hat E}^{\beta}\right]=N^{\alpha\beta}
{\hat E}^{\alpha+\beta}.
\end{eqnarray}
Roots $\alpha(i)$ and parameters $\alpha^i,\; N^{\alpha\beta}$
are structure constants of the algebra in the Cartan -- Weyl basis.
Our goal is to construct nilpotent BRST charge, which after quantization
 leads to the following conditions: all positive root vectors
${\hat E}^\alpha\;(\alpha>0)$ of
the algebra annihilate the physical states. Contrary, the operators
${\hat H}^i$ which form the Cartan subalgebra
may or may not be constraints, depending
on the physical nature of these operators.

The simplest case, when
all Cartan generators annihilate the physical states, is well known.
We introduce the set of anticommuting variables
$\e_i,\e_\alpha,$ $\e_{-\alpha}=\e_\alpha^+$, having ghost number one
and corresponding momenta
$\cp_i,\cp_{-\alpha}=\cp_\alpha^+,\cp_\alpha$, with the
commutation relations:
\begin{equation}
\{\e_i,\cp_k\}=\delta_{ik},\;\{\e_\alpha,\cp_{-\beta}\}=
\{\e_{-\alpha},\cp_\beta\}=\delta_{\alpha\beta}
\end{equation}
we define the ``ghost vacuum" as
\begin{equation}
\e_\alpha|0\rangle=\cp_\alpha|0\rangle =\cp_i|0\rangle=0
\end{equation}
for positive roots $\alpha$.
The BRST charge for the Cartan -- Weyl decomposition of the algebra
has a standard form
\begin{eqnarray}
Q&=&\sum_i\e_i {\hat H}^i+\sum_{\alpha>0}\left(\e_\alpha {\hat E}^{-\alpha}+
\e_{-\alpha}{\hat E}^{\alpha}\right)-
\frac{1}{2}\sum_{\alpha\beta}N^{\alpha\beta}
\e_{-\alpha}\e_{-\beta}\cp_{\alpha+\beta}
+        \nonumber      \\                    \label{brst1}
&&\sum_{\alpha>0,i}\{\alpha(i)\left(\e_i\e_\alpha\cp_{-\alpha}-
\e_i\e_{-\alpha}\cp_\alpha\right)+\alpha^i\e_\alpha\e_{-\alpha}\cp_i
\}
\end{eqnarray}
The physical states are then the cohomology classes of the BRST operator.

The quantization in this case is similar to the quantization
${\grave a}$ la Gupta -- Bleuler, because physical states satisfy
equations $$({\hat H}^i + \sum_{\alpha>0}\alpha(i)) |Phys\rangle=0,\quad 
{\hat E}^\alpha|Phys\rangle=0$$
only for positive values of $\alpha$. The appearance of the
 $\sum\alpha(i)$ in the quantization conditions
does not cause  problems  since these terms can be absorbed after 
the redefinition of $({\hat H}^i$ as we shall see below.

The situation becomes different if some of the Cartan operators
${\hat H}^i$, say ${\hat H}^{i_l},\;l=1,2,...N$  are
nonvanishing from the  physical reasons. In this case the following
method can be used.

First of all we construct some auxiliary representation for the
generators ${\hat H}^i,\; {\hat E}^\alpha$ of the
algebra in terms of additional creation and annihilation operators.
The only condition for this representation is that it depends
on some parameters $h^{n}$. The total number of
these parameters is equal to  the
number of the Cartan generators, which are nonzero in the physical
sector. In what follows, we  consider the realizations of the
algebra with a linear dependence of the Cartan generators on these
parameters: $H^{m}(h)=\tilde{H}^{m}+c^m_n h^{n}$, where
$c^m_n$ are  some constants.
The  $h^n$ dependence of other generators can be arbitrary.
In the next section
we describe the general method of construction of such representations.
Here we simply assume that they exist.

The next step is to consider the realization of the algebra
as a sum of "old" and "new" generators
\begin{equation} \label{sum}
{\cal H}^i={\hat H}^i + {H}^i(h) ,\quad
{\cal E}^{\alpha}={\hat E}^{ \alpha} + E^{ \alpha}(h).
\end{equation}
The BRST charge for the total system has the same form as
\p{brst1}, with modified generators:
\begin{eqnarray}
{\cal Q}&=&\sum_i\e_i {\cal H}^i+
\sum_{\alpha>0}\left(\e_\alpha {\cal E}^{-\alpha}+
\e_{-\alpha}{\cal E}^{\alpha}\right)-
\frac{1}{2}\sum_{\alpha\beta}N^{\alpha\beta}
\e_{-\alpha}\e_{-\beta}\cp_{\alpha+\beta}
+        \nonumber      \\                    \label{brst2}
&&\sum_{\alpha>0,i}\{\alpha(i)\left(\e_i\e_\alpha\cp_{-\alpha}-
\e_i\e_{-\alpha}\cp_\alpha\right)+\alpha^i\e_\alpha\e_{-\alpha}\cp_i
\}
\end{eqnarray}
The ghost variables $\eta_{i_l}$, correspond  to the set of
nonvanishing generators ${\hat H}^{i_l}$ and therefore one needs to
remove the $\eta_{i_l}$ dependence
\begin{equation}
{\cal Q}_{i_l}=\e_{i_l}\{
{\hat H}^{i_l} +\tilde{H}^{i_l}+c^{i_l}_nh^n
+\sum_{\beta>0}\beta(i_l)\left(\e_\beta\cp_{-\beta}-
\e_{-\beta}\cp_\beta\right)
\}.
\end{equation}
from the BRST  charge. For this purpose
consider an auxiliary $N$ - dimensional
space with coordinates $x_{i_l}$ and conjugated momenta $p^{i_l}$,
where $c^{i_l}_nh^n=p^{i_l}$:
\begin{equation}
\left[x_{i_l}, p^{i_n}\right]=i \delta_{i_l}^{i_n}.
\end{equation}
After the similarity transformation, which corresponds
to the dimensional reduction \cite{PT2}
\begin{equation}       \label{transformation}
\tilde{\cal Q}=e^{i\pi^{i_l} x_{i_l}} {\cal Q} e^{-i\pi^{i_l} x_{i_l} },
\end{equation}
where
\begin{equation}
\pi^{i_l}=
{\hat H}^{i_l} +\tilde{H}^{i_l}+
\sum_{\beta>0}\beta(i_l)\left(\e_\beta\cp_{-\beta}-
\e_{-\beta}\cp_\beta\right)
\end{equation}
the transformed BRST charge $\tilde{\cal Q}$ does not depend on the ghost
variables $\eta_{i_l}$. All parameters $p^{i_l}$ in the BRST charge
are replaced by the corresponding operators $-\pi^{i_l}$.
The transformation \p{transformation}
does not change the nilpotency property of the BRST charge.
It means that the $\cp_{i_l}$ independent part $\tilde{\cal Q}_0$ of the
total charge $\tilde{\cal Q}$ is nilpotent as well.
Moreover, as a consequence of the nilpotency of $\tilde{\cal Q}$
all coefficients  at the corresponding
antighost operators  $\cp_{i_l}$ commute with $\tilde{\cal Q}_0$.
One can show that the quantization with the help of the BRST
operator $\tilde{\cal Q}_0$ will lead to the desirable reduced system of
constraints on the physical states.

\setcounter{equation}0\section{Construction of auxiliary representations
of the algebra}
Consider the highest weight representation of the algebra under
consideration with the highest weight vector ${|\Phi\rangle}_V$, annihilated
by the positive roots
\begin{equation}\label{V1}
{E^\alpha|\Phi\rangle}_V=0
\end{equation}
and being the proper vector of the Cartan generators
\begin{equation}\label{V2}
H^i{|\Phi\rangle}_V=h^i{|\Phi\rangle}_V.
\end{equation}
As it was  shown  in \cite{PTPR} the representations of this algebra
can be (in principle) constructed by means of 
the so called Gelfand -- Tsetlin schemes  \cite{GT}. However difficulties 
in such construction arise, if one considers algebras,
different from the simplest ones of rank $1$.

In this section we describe another method, based on the construction
 given in \cite{B}.
The representation which is given by \p{V1} and \p{V2}
 in the mathematical literature is called  the Verma
module \cite{D}. Following the Poincare -- Birkhoff -- Witt theorem,
 the basis space of this representation
is given by vectors
\begin{equation}\label{b1}
{\left|n_1,n_2,\dots,n_r \right \rangle}_V =
(E^{-\alpha_1})^{n_1}(E^{-\alpha_2})^{n_2}\dots (E^{-\alpha_r})^{n_r}{|\Phi \rangle}_V
\end{equation}
where ${\alpha_1}, {\alpha_2}, \dots {\alpha_r}$ is some ordering of positive roots
 and $n_i \in N$.

 Using the commutation relations of the  algebra and the formula 
$$
AB^n=\sum^n_{k-1}{n \choose k }B^{n-k}[[\dots [A,B],B] \dots ]
$$
one can calculate the explicit form of the Verma module.
 In  \cite{B} it was shown  that,  making use of the map 
\begin{equation}\label{cb2}
{\left|n_1,n_2,\dots,n_r \right \rangle}_V  
\longleftrightarrow \left|n_1,n_2,\dots,n_r \right\rangle
\end{equation}
where $\left|n_1,n_2,\dots,n_r \right\rangle$   are 
 base vectors of the Fock space 
  \begin{equation}\label{b2}
\left|n_1,n_2,\dots,n_r \right\rangle =
(b^+_1)^{n_1}(b^+_2)^{n_2}\dots (b^+_r)^{n_r}|0 \rangle.
\end{equation}
generated by creation and annihilation operators 
$b_i^+, b_i \; i=1,2,\dots,r$ with the standard commutation relations
\begin{equation}\label{heisenberg}
\left[b_i,b_j^+ \right] =\delta_{ij},
\end{equation}
the Verma module can be rewritten as polynomials in 
creation operators on  the Fock space.

As an explicit example of the construction  given above
let us consider the representations
of  $so(3,2)$ algebra, which can be used for the 
description of the higher spin fields.
In this case  commutation relations between the corresponding
generators $L_1^+, L_{12}^+, L_2^+, T^+, L_1, L_{12}, L_2, T, H_1, H_2 $ 
 are given  by
\begin{eqnarray}
&&[L_1,L_1^+]=H_1, \quad  [L_2,L_2^+]=H_1, \quad [L_{12},L_{12}^+]=H_1+H_2, \nonumber \\
&&[T^+,L_1^+]=-L^+_{12}, \quad  [T^+,L^+_{12}]=-2L^+_2, \quad  \quad [T,L_1^+]=0, \nonumber \\
&&[T,L_1]=L_{12}, \quad  [T,L_{12}]=-2L_2, \quad  [T,T^+]=H_1-H_2, \nonumber \\
&&[L_2,L_{12}^+]=-T, \quad  [L_2,T^+]=-L_{12}, \quad  [T,L_2^+]=-L^+_{12}, \\
&&[L_1,L_{12}^+]=-T^+, \quad  [L_{12},L_1^+]=-T, \quad  [L_{12},L^+_2]=-T^+, \nonumber \\
&&[T,L_{12}]=-2L_{1}^+, \quad  [T^+,L_{12}]=2L_1, \nonumber \
\end{eqnarray}
In this case we take the representation space the following 
space of vectors
$$
{\left|n \right>}_V={\left|n_1, n_2, n_3, n_4 \right>}_V=
(L_1^+)^{n_1}(L_{12}^+)^{n_2}(L_2^+)^{n_3}(T^+)^{n_4}{|\Phi\rangle}_V
$$
and after the simple calculations  one obtains
\begin{eqnarray}\label{VER}
&&L_1^+{\left|n \right>}_V={\left|n_1+1, n_2, n_3, n_4 \right>}_V , \; 
L_{12}^+{\left|n \right>}_V={\left|n_1, n_2+1, n_3, n_4 \right>}_V , \nonumber \\
&&L_2^+{\left|n\right>}_V={\left|n_1, n_2, n_3+1, n_4 \right>}_V, \nonumber \\
&&T^+{\left|n\right>}_V={\left|n_1, n_2, n_3, n_4 +1 \right>}_V-
n_1{\left|n_1-1, n_2+1, n_3, n_4 \right>}_V\nonumber \\
&&\hskip0.5cm -2n_2{\left|n_1, n_2-1, n_3+1, n_4 \right>}_V, \nonumber \\
&& H_1{\left|n \right>}_V=(2n_1+n_2-n_4+h_1){\left|n\right>}_V, \nonumber \\ 
&&H_2{\left|n \right>}_V=(n_2+2n_3+n_4+h_2){\left|n\right>}_V,  \\
&&L_1{\left|n \right>}_V=(n_1+n_2-n_4+h_1-1)n_1{\left|n_1-1, n_2, n_3, n_4 \right>}_V \nonumber \\
&&\hskip0.5cm -n_2{\left|n_1, n_2-1, n_3, n_4+1 \right>}_V+n_2(n_2-1)
{\left|n_1, n_2-2, n_3+1, n_4 \right>}_V,\nonumber \\
&&L_{12}{\left|n \right>}_V= (2n_1+n_2+2n_3+h_1+h_2-1)n_2
{\left|n_1, n_2-1, n_3, n_4 \right>}_V \nonumber \\
&&\hskip0.5cm - n_3{\left|n_1, n_2, n_3-1, n_4+1 \right>}_V+ n_3n_1
{\left|n_1-1, n_2+1, n_3-1, n_4 \right>}_V \nonumber \\
&&\hskip0.5cm +n_4(n_4-1)n_1{\left|n_1-1, n_2, n_3, n_4-1 \right>}_V \nonumber \\
&&\hskip0.5cm +(h_2-h_1)n_4n_1{\left|n_1-1, n_2, n_3, n_4-1 \right>}_V ,\nonumber \\
&&L_{2}{\left|n \right>}_V= (n_2+n_3+n_4+
h_2-1)n_3{\left|n_1, n_2, n_3-1, n_4 \right>}_V \nonumber \\ 
&&\hskip0.5cm +n_2(n_2-1){\left|n_1+1, n_2-2 n_3, n_4 \right>}_V \nonumber \\
&&\hskip0.5cm +(h_2-h_1+n_4-1)n_4n_2{\left|n_1, n_2-1, n_3, n_4-1 \right>}_V, \nonumber \\
&&T{\left|n\right>}_V=(h_1-h_2-n_4+1)n_4{\left|n_1, n_2, n_3, n_4 -1 \right>}_V \nonumber \\
&&\hskip0.5cm -2n_2{\left|n_1+1, n_2-1, n_3, n_4 \right>}_V
 -n_3{\left|n_1, n_2+1, n_3-1, n_4 \right>}_V. \nonumber \
\end{eqnarray}
It can be seen that the action of the  following operators  in the 
Fock space 
\begin{eqnarray}
&&L_1^+=b_1^+, \quad L_{12}^+=b_2^+, \quad L_2^+= b_3^+, \nonumber \\
&&T^+=b_4^+-b_2^+b_1-2b_3^+b_2, \nonumber \\
&&T=(h_1-h_2-b_4^+b_4)b_4-2b_1^+b_2-b_2^+b_3, \nonumber \\
&&H_1=2b_1^+b_1+b_2^+b_2-b_4^+b_4+h_1 \nonumber, \\
&&H_2=b_2^+b_2+2b_3^+b_3+b^+_4b_4+h_2, \\
&&L_1=(b_1^+b_1+b_2^+b_2-b_4^+b_4+h_1)b_1-b_4^+b_2+b_3^+b_2b_2, \nonumber \\
&&L_{12}=(2b_1^+b_1+b_2^+b_2+2b_3^+b_3+h_1+h_2)b_2 + b^+_4b_4b_4b_1 \nonumber \\
&&\hskip1cm +b_2^+b_3b_1-b_4^+b_3 + (h_2-h_1)b_4b_1, \nonumber \\
&&L_{2}=(b_2^+b_2+b_3^+b_3+b_4^+b_4+h_2)b_3 \nonumber \\ 
&&\hskip1cm +(h_2-h_1)b_4b_2 
+b_1^+b_2b_2+b_4^+b_4b_4b_2. \nonumber \
\end{eqnarray}
is identical to the expressions \p{VER} for the Verma module.

As it  can be concluded from the realization of the $so(3,2)$ algebra,
the generators, corresponding to opposite roots 
are not hermitian conjugated
to each other. Indeed, following the usual rules of hermitian conjugation
for creation and annihilation operators
\begin{equation} \label{osc}
(b_i)^+=b_i^+,\quad (b_i^+)^+=b_i        
\end{equation}
the  operator conjugated to $L_1^+$   is not equal to the operator
$L_1$ since $(L_1^+)^+=(b_1^+)^+=b_1 \neq L_1 $.
 The same statement
is true for all other pairs of root generators. It means that the 
BRST charge ${\cal Q}$, constructed with the help of these operators,
 though being nilpotent,
is not hermitian.  This causes the serious  problems, because
the BRST gauge invariance has been lost and consequently the 
lagrangians of the form $L\sim \langle\Psi|{\cal Q}|\Psi\rangle$
\cite{PT2}, \cite{SI} -- \cite{PT3} are no longer gauge invariant.

\setcounter{equation}0\section{The restoration of the hermiticity properties}

The situation, when the generators corresponding to the opposite
roots are not mutually hermitian conjugated holds for any
 algebra under consideration and is a consequence of the 
method used  for the  construction of the 
corresponding representations. 

The reason is that
if we consider the usual scalar product in the Fock space with basis
\p{b2}, we  find that these vectors  form the orthogonal (not 
orthonormal) basis. At the same time the corresponding vectors \p{b1}
in Verma module are not orthogonal. For example, the scalar product
of two vectors 
\begin{equation} \label{example12}
|\Phi_1\rangle_V= L_1^+T^+|\Phi\rangle_V, \quad |\Phi_2\rangle_V=L_{12}^+|\Phi\rangle_V
\end{equation}
is different from zero
\begin{equation} \label{example12a}
{}_V\langle\Phi_1 |\Phi_2\rangle_V={}_V\langle\Phi|T L_1L_{12}^+|\Phi\rangle_V=(h_2-h_1),
\end{equation}
where we assumed that ${}_V\langle\Phi|\Phi\rangle_V=1$.
Therefore the correspondence between these two spaces 
is not complete because of the difference in 
the scalar products of pairs of corresponding vectors.

The idea how to improve the situation lies on the modification of the scalar 
product in the auxiliary Fock space.
The standard scalar product of two vectors of type 
\p{b2}, namely $|\Phi_1\rangle$ and $|\Phi_2\rangle$ is
defined as 
\begin{equation}\label{st}
(\Phi_1 |\Phi_2)_{st}=\langle\Phi_1 |\Phi_2\rangle
\end{equation}
and is calculated by transition of the annihilation 
operators $b_i$ to the right
by means of the commutation relations \p{heisenberg} 
with the subsequent use of
 the property $b_i|0\rangle = 0$.

Let us introduce  the new scalar product in the Fock space
\begin{equation} \label{new}
(\Phi_1 |\Phi_2)_{new}=\langle\Phi_1 |K|\Phi_2\rangle,
\end{equation}
with a kernel $K$, which depends on 
the creation and annihilation operators
$b_i$ and $b_i^+$. The only condition on this 
new scalar product in the Fock space
is that it has to coincide  with the scalar product in the Verma module.
Therefore  
taking two arbitrary vectors in the Verma module 
$|\Phi_1\rangle_V, |\Phi_2\rangle_V$
and corresponding vectors in the Fock space $|\Phi_1\rangle, |\Phi_2\rangle$
one has the following defining relation:
\begin{equation} \label{def}
\langle\Phi_1 |K|\Phi_2\rangle={}_V\langle\Phi_1 |\Phi_2\rangle_V.
\end{equation}
According to this relation  the hermiticity properties of  the root
generators are restored in the following sense.
Let us consider the scalar product of the states
$|\Phi_1\rangle_V$ and $ E^\alpha|\Phi_2\rangle_V$: 
${}_V\langle\Phi_1 |E^\alpha|\Phi_2\rangle_V$. Due to hermitian
properties of the root generators in the Verma module it coincides 
with the scalar product of the states 
$E^{-\alpha}|\Phi_1\rangle_V$ and $ |\Phi_2\rangle_V$ since
\begin{equation} \label{con}
(E^{-\alpha}|\Phi_1\rangle_V)^+={}_V\langle\Phi_1 |E^\alpha.
\end{equation}
In the Fock space the relation \p{con} looks as
\begin{equation} \label{confock}
(E^{-\alpha}|\Phi_1\rangle_V)^+={}_V\langle\Phi_1 |(E^{-\alpha})^+.
\end{equation}
So, taking the new scalar product of
the pairs of corresponding vectors in the Fock space
$|\Phi_1\rangle, E^\alpha|\Phi_2\rangle$ and
$E^{-\alpha}|\Phi_1\rangle, |\Phi_2\rangle$ 
one has the following relations
\begin{equation} \label{reality}
\langle\Phi_1 |KE^\alpha|\Phi_2\rangle=
 \langle\Phi_1 |(E^{-\alpha})^+K|\Phi_2\rangle.
\end{equation}
Therefore all of the root generators of the algebra 
under consideration satisfy the  relations
\begin{equation} \label{hermiticity}
KE^\alpha=(E^{-\alpha})^+K,
\end{equation}
which play the role of hermiticity relations.

Now, consider the part of the BRST charge in the Fock space dependent 
on the root generators
\begin{equation} {\cal Q}_{nonh}=
\sum_{\alpha>0}\left(\e_\alpha E^{-\alpha}+
\e_{-\alpha}E^{\alpha}\right).
\end{equation}
being the only non -- hermitian part of the BRST charge
\begin{equation} {\cal Q}_{nonh}^+=
\sum_{\alpha>0}\left(\e_{-\alpha} (E^{-\alpha})^++
\e_{\alpha}(E^{\alpha)^+}\right)\neq {\cal Q}_{nonh}.
\end{equation}
 It can be easily shown, that the following relations take place
$${\cal Q}_{nonh}^+K=K{\cal Q}_{nonh},\quad 
{\cal Q}_{nonh}K=K{\cal Q}_{nonh}^+$$
So one can conclude, that the total BRST charge of the form \p{brst2}
constructed with the help of the generators  \p{sum}
satisfies the modified hermiticity relation
\begin{equation}\label{hr}
{\cal Q}^+K=K{\cal Q}.
\end{equation}
This gives the possibility to construct the lagrangians of the form
$L\sim \langle\Psi|K{\cal Q}|\Psi\rangle$, which  are gauge invariant
with the following transformation rules for the field $|\Psi\rangle$ 
\begin{equation}
\delta|\Psi\rangle={\cal Q}|\Psi\rangle,\quad 
\delta \langle\Psi|=\langle\Psi|{\cal Q}^+.
\end{equation}
Obviously the gauge invariance is guaranteed by the 
nilpotency of ${\cal Q}$ and ${\cal Q}^+$
and by the relation \p{hr}.

Bellow we  prove the existence of the hermitian kernel $K$ and show,
how it can be constructed. The central role in this construction will
play the matrix of scalar products of basic elements of the Verma module
\begin{equation}
C_{n_1,\cdots,n_r}^{m_1,\cdots,m_r}\equiv {}_V\langle n_1,\cdots,n_r|
m_1,\cdots,m_r\rangle_V.
\end{equation}
~Let ~us ~introduce ~the ~notion ~of ~ancestor ~for ~the 
~pair ~of ~multiindeces 
$\{n_1,\cdots,n_r|
m_1,\cdots,m_r\}$. 
~It ~is ~defined ~in ~the ~following ~way.
Consider a pair of indices $\{n_k|m_k\}$ standing on the $k$ -th place 
of the given pair of multiindices. The ancestor is the pair of representations,
which has on the $k$ -th place the following pair:

$\{n_k-m_k|0\}$ if $n_k>m_k$;

$\{0|m_k-n_k\}$ if $n_k<m_k$;

$\{0|0\}$ if $n_k=m_k$;\\
It means that we reduce the pair on the maximal common number.
We can illustrate graphically this procedure for the case of $SO(3,2)$ algebra,
which has the rank $2$. Its root diagram is shown on the following picture.

\vspace*{6cm}

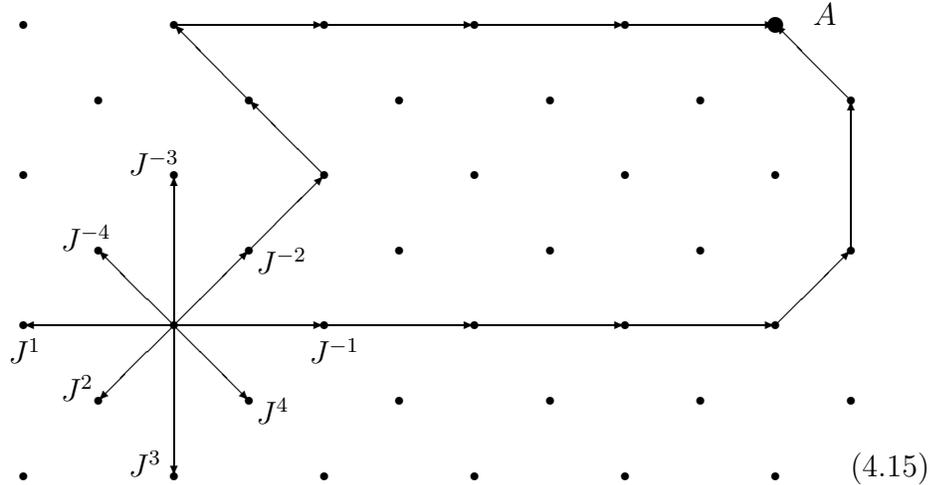
\begin{equation} \nonumber
\begin{picture}(11,6)
\unitlength=1cm
\multiput(-5,6)(2,0){6}{\circle*{.1}}
\multiput(-5,0)(2,0){6}{\circle*{.1}}
\multiput(-5,2)(2,0){6}{\circle*{.1}}
\multiput(-5,4)(2,0){6}{\circle*{.1}}
\multiput(-4,1)(2,0){6}{\circle*{.1}}
\multiput(-4,3)(2,0){6}{\circle*{.1}}
\multiput(-4,5)(2,0){6}{\circle*{.1}}
\put(-3,2){\vector(1,0){2}}
\put(-3,2){\vector(-1,0){2}}
\put(-3,2){\vector(0,1){2}}
\put(-3,2){\vector(0,-1){2}}
\put(-3,2){\vector(1,1){1}}
\put(-3,2){\vector(1,-1){1}}
\put(-3,2){\vector(-1,1){1}}
\put(-3,2){\vector(-1,-1){1}}
\put(-1,2){\vector(1,0){2}}
\put(1,2){\vector(1,0){2}}
\put(3,2){\vector(1,0){2}}
\put(-3,6){\vector(1,0){2}}
\put(-1,6){\vector(1,0){2}}
\put(1,6){\vector(1,0){2}}
\put(3,6){\vector(1,0){2}}
\put(-2,3){\vector(1,1){1}}
\put(5,2){\vector(1,1){1}}
\put(-1,4){\vector(-1,1){1}}
\put(-2,5){\vector(-1,1){1}}
\put(6,5){\vector(-1,1){1}}
\put(6,3){\vector(0,1){2}}
\put(5,6){\circle*{.2}}
\put(5.5,6){$A$}
\put(-5.2,1.5){$J^{1}$}
\put(-1.2,1.5){$J^{-1}$}
\put(-3.6,0){$J^{3}$}
\put(-3.6,4){$J^{-3}$}
\put(-4.5,1){$J^{2}$}
\put(-1.9,0.7){$J^{4}$}
\put(-4.5,3){$J^{-4}$}
\put(-1.9,2.7){$J^{-2}$}
\end{picture}
\label{path}
\end{equation}

 Here are drown two different vectors of the Verma module.
Each vector corresponds to the line, which begins at the origin and 
ends at the point $A$.
Different segments of these lines correspond to negative roots, which 
are present in the definition of the vector $|n_1,\cdots,n_r\rangle_V$.
The first (lower)  line corresponds to the vector $|4,1,1,1\rangle_V$,
while the second one corresponds to the vector $|4,2,0,2\rangle_V$.
The lines described above are not unique for the given vectors, since 
all lines with the same numbers of each negative roots
represent the same vector.  However this fact does not affect the result
obtained with the help of this picture.
The vectors with corresponding lines ended at the same point, say
at the point $A$, are the only ones which
can have nonzero scalar product.
The pair of representations for the vectors drawn on the picture is 
$\{4,1,1,1|4,2,0,2\}$, while the corresponding 
ancestor is $\{0,0,1,0|0,1,0,1\}$.
In general all pairs of representations are divided into the equivalence classes
by their ancestors.

The following expression solves the problem of finding the kernel $K$.
\begin{equation}\label{kernel}
K=\sum_{anc}    (b_1^+)^{n_1}\cdots (b_r^+)^{n_r}\;     
C_{n_1,\cdots,n_r}^{m_1,\cdots,m_r}(b^+_1b_1,b_2^+b_2,\cdots,b_r^+b_r)\;(b_1)^{m_1}
\cdots (b_r)^{m_r},
\end{equation}
~where  ~the ~summation ~goes ~over ~all ~possible ~ancestors ~and \\
~$C_{n_1,\cdots,n_r}^{m_1,\cdots,m_r}(b^+_1b_1,b_2^+b_2,\cdots,b_r^+b_r)$ 
~are
~the ~functions ~of ~the ~number ~operators 
~$b^+_1b_1,b_2^+b_2,\cdots,b_r^+b_r$
~with ~the ~following ~properties:
\begin{equation}
C_{n_1,\cdots,n_r}^{m_1,\cdots,m_r}(0,0,\cdots,0)=
C_{n_1,\cdots,n_r}^{m_1,\cdots,m_r},
\end{equation}
\begin{equation}
C_{n_1,\cdots,n_r}^{m_1,\cdots,m_r}(l_1,\cdots,l_r)=
C_{n_1+l_1,\cdots,n_r+l_r}^{m_1+l_1,\cdots,m_r+l_r},
\end{equation}
The hermiticity of the kernel $K$ is a consequence of the relations
\begin{equation}
C_{n_1,\cdots,n_r}^{m_1,\cdots,m_r}(l_1,\cdots,l_r)=
C^{n_1,\cdots,n_r}_{m_1,\cdots,m_r}(l_1,\cdots,l_r)
\end{equation}

Therefore, the expression \p{kernel} solves the problem of finding the kernel
for the scalar product for an arbitrary Lie algebra.
\setcounter{equation}0\section{Conclusions}
It might be interesting to apply the method
described in the paper to construct the gauge invariant
 lagrangians for the particles with the higher spins \cite{LA1} --
\cite{LA2}. Namely, the case of $so(3,2)$ algebra which we have considered
in details corresponds to the subset of constraints obtained after the
quantization of the three -- particle bound system (three -- point discrete
string) \cite{GP}. The description of the various irreducible
representations of the Poincare  group with the corresponding
Young tableau having two rows  can be achieved after elimination 
of the Cartan generators of $so(3,2)$ algebra
from the total set of constraints.
The application of the BRST approach given in this letter to the description
of the above mentioned system will be given elsewhere.
\vspace{.5cm}\\

\noindent {\bf Acknowledgments.}
This investigation has been supported in part by the
Russian Foundation of Fundamental Research,
grant 99-02-18417,
joint grant RFFR-DFG 99-02-04022, grant INTAS 96-0308 and
grant of the Committee for collaboration Czech Republic with JINR.
One of us (M.T.) is grateful to the Abdus Salam International
Centre for Theoretical Physics, Trieste, 
where the part of the present work was done.


\begin{thebibliography}{99}
\bibitem{FV} E.  S. Fradkin, G. A. Vilkovisky.  Phys.Lett. 
{\bf B55},  224 (1975).
\bibitem{BV} I. A. Batalin, G. A. Vilkovisky.  Phys.Lett. 
{\bf B69},  309 (1977).
\bibitem{FS} L. D. Faddeev, S. L. Shatashvili. Phys.Lett. 
{\bf B167},  225 (1986).
\bibitem{BF2} I. A. Batalin, E. S. Fradkin. Nucl.Phys. 
{\bf B279},  514 (1987).
\bibitem{EM}  E. T. Egoryan, R. P. Manvelyan. Theor. Math.Phys.
{\bf 94},  241 (1993).
\bibitem{PT2} A. Pashnev, M. Tsulaia. Mod.Phys.Lett. 
{\bf A12},  861 (1997).
\bibitem{PTPR} A. Pashnev, M. Tsulaia. Proceedings 
of the International
conference {\it ``Supersymmetry and Quantum Symmetries"}  
Springer  348 (1999).
\bibitem{GT} I. M. Gel'fand, M. L. Tsetlin. Dokl. Akad. Nauk SSSR.
{\bf71},  825 (1950); ibid {\bf71},  1017 (1950).
\bibitem{B}{\v{C}. Burd\'{\i}k:} J.Phys.A: Math.Gen. 
{\bf 18},  3101 (1985).
\bibitem{D}{J. Dixmier, }{\it Algebres enveloppantes},  
Gauthier-Villars, Paris (1974).
\bibitem{SI} W. Siegel.  Phys.Lett. 
{\bf B149},  157 (1985).
\bibitem{OS} S.Ouvry, J.Stern. Phys.Lett. {\bf B177},  335 (1986).
\bibitem{PT3} A. Pashnev, M. Tsulaia. Mod.Phys.Lett. 
{\bf A13},  1853 (1998).
\bibitem{LA1} J. M. F. Labastida. Phys.Rev.Lett. {\bf 58},  531 (1987).
\bibitem{LA2} J. M. F. Labastida. Nucl.Phys. {\bf B322},  185 (1989).
\bibitem{GP} V. D. Gershun, A. I. Pashnev. Theor. Math. Phys.
{\bf 73}, 294 (1987).
\end{thebibliography}
\end{document}